# Countries pushing the boundaries of knowledge: the USA's dominance, China's rise, and the EU's stagnation

Alonso Rodríguez-Navarro

*Departamento de Biotecnología-Biología Vegetal, Universidad Politécnica de Madrid, 28040, Madrid, Spain*
*Departamento de Estructura de la Materia, Física Térmica y Electrónica y GISC, Universidad Complutense de Madrid, 28040, Madrid, Spain*

*E-mail address: alonso.rodriguez@upm.es*

Knowing which countries contribute the most to pushing the boundaries of knowledge in science and technology has social and political importance. However, common citation metrics do not adequately measure this contribution. This measure requires more stringent metrics appropriate for the highly influential breakthrough papers that push the boundaries of knowledge, which are very highly cited but very rare. Here I used the recently described $Rk$-index, specifically designed to address this issue. I applied this index to 25 countries and the EU across 10 key research topics—five technological and five biomedical—studying domestic and international collaborative papers independently. In technological topics, the $Rk$-indices of domestic papers show that overall, the USA, China, and the EU are leaders; other countries are clearly behind. The USA is notably ahead of China, and the EU is far behind China. The same approach to biomedical topics shows an overwhelming dominance of the USA and that the EU is ahead of China. The analysis of internationally collaborative papers further demonstrates the USA's dominance. These results conflict with current country rankings based on less stringent indicators.

*Keywords*: Citation analysis, Scientific progress, Highly cited, Rank analysis, Scientometrics

## 1. Introduction

Research investments are quite high in developed countries and involve a significant economic effort in most countries. In this global context, society needs to understand what it receives in return, a task that is difficult to fulfill because the product of research, knowledge, cannot be easily measured. Furthermore, because "the benefits of scientific discovery have been heavy-tailed" (Press, 2013), the assessment of these discoveries that push the boundaries of knowledge has added difficulties because of their low frequency. Therefore, although experts have warned for many years that "Government policy-makers, corporate research managers, and university administrators need valid and reliable S&T indicators" (Garfield & Welljams-Dorof, 1992) and that "The ability to judge a nation's scientific standing is vital for the governments, businesses and trusts" (King, 2004), a definitive solution to meet these requirements is still pending.

Citation metrics accurately gauge the scientific impact of research (Aksnes et al., 2019; Waltman, 2016) when used at high aggregation levels (van Raan, 2005). Percentile indicators are considered the most reliable of these metrics (Bornmann & Marx, 2013). However, they do not assess the contribution to pushing the boundaries of knowledge, despite this contribution is one of the most important goals of research. This failure occurs because, while the boundaries of knowledge are pushed by very rare breakthrough publications, current metrics are based on relatively common publications. The most stringent rankings normally use the number of top 1% most cited papers (Bornmann, 2014), but breakthrough papers account for only about 0.01% of all published papers (Bornmann et al., 2018; Poege et al., 2019). A second difficulty arises because in technologically advanced countries, research has two aims: to push the boundaries of knowledge and to boost incremental innovation in technological industries. Citation distributions in these two types of research are different (Rodríguez-Navarro & Brito, 2022), and a single comprehensive citation metric for both together cannot be obtained.

The European Commission (European Commission, 2022), the US National Science Board (National Science Board, 2022), the Organization for Economic Cooperation and Development (OECD, 2017), and other institutions use the number of top 10% or 1% most cited papers to produce country rankings. At this mild stringent level, these rankings do not accurately reflect the contribution to pushing the boundaries of knowledge. This goal is not normally explicitly stated in these rankings, but they convey a notion of excellence that policy makers and other readers may incorrectly identify with the capacity to achieve breakthroughs. The poor research evaluations of Japan are the most notable failure of these metrics (Bornmann & Leydesdorff, 2013; Pendlebury, 2020; Rodríguez-Navarro & Brito, 2024a).

Another complication arises from international collaborations, which are very common. Current metrics analyze these collaborations fractioning the merits without considering potential unequal distribution among collaborative countries (Olechnicka et al., 2019; Zanotto et al., 2016).

## 1.1. The Rk-index

The primary challenge of evaluating countries based on their contribution to pushing the boundaries of knowledge lies in the infrequency of such contributions. This requires specific metrics, and the *Rk*-index was developed with this purpose (Rodríguez-Navarro & Brito, 2024b). The well-known percentile indicators count the number of a country's papers in a certain top percentile of the global list ordered by the number of citations (Bornmann & Marx, 2013), and in this aggregation, breakthroughs are counted together with less cited papers. In contrast, the *Rk*-index utilizes the global ranks of the 10 most cited papers, and is calculated by summing 20 to these ranks and taking the geometric mean. Table 1 presents the ranks of the 10 most cited domestic papers from the USA, China, and the EU in the topic of solar cells/photovoltaics from a total of 61,699 papers. These series of 10 numbers intuitively reveal the dominance of the USA. This is also suggested by the top percentiles approach, but with poor statistical support. In this approach, of the 62 most cited papers in the world (top 0.1%), the USA has 10, China has four and the EU has two (papers in Table 1 with ranks below 62). At a more stringent level, among the six most cited papers in the world (top 0.01%), the USA has one, and China and the EU have zero (papers in Table 1 with ranks below 7). These low figures make it clear that percentile indicators cannot be used to rank countries by their contribution to pushing the boundaries of knowledge. In contrast, the *Rk*-index provides robust results that are equivalent to the number of the top 0.01% most cited papers (Rodríguez-Navarro & Brito, 2024b).

**Table 1.** Global ranks of the most cited domestic papers in the topic of solar cells/photovoltaics and corresponding *Rk*-indices

| Local ranks | Global ranks | | |
|---|---|---|---|
| | **USA** | **China** | **EU** |
| 1 | 4 | 10 | 46 |
| 2 | 8 | 18 | 48 |
| 3 | 9 | 19 | 95 |
| 4 | 15 | 22 | 118 |
| 5 | 24 | 77 | 121 |
| 6 | 27 | 97 | 133 |
| 7 | 29 | 106 | 141 |
| 8 | 31 | 108 | 167 |
| 9 | 32 | 109 | 203 |
| 10 | 36 | 132 | 209 |
| ***Rk*-index** | | | |
| 39.47 | 25.05 | 13.10 | 7.29 |

Publication window 2014–2017, citation window 2019–2022. The *Rk*-index of local ranks is included to show its maximum value

## 1.2. Aim of this study

To calculate the contribution to pushing the boundaries of knowledge, I computed the *Rk*-index for 25 countries plus the EU in fast-evolving research topics: five technological and five biomedical topics that currently hold central positions in world research. This focused approach, as utilized by the *Australian Strategic Policy Institute* (Gaida et al., 2023), is advantageous compared to analyzing the totality of all research or wide research fields. Differences between topics can be significant, and averaging over numerous topics may obscure important information. For instance, I show below that China excels in graphene research but lags considerably behind the EU in Alzheimer's and Parkinson's diseases. Moreover, I independently calculated the *Rk*-indices for both domestic and international collaborative papers due to the previously mentioned reasons of potential unequal distribution among collaborative countries (Olechnicka et al., 2019; Zanotto et al., 2016).

## 2. Methods

Searches were performed in the Science Citation Index Expanded database in the Clarivate's Web of Science Core Collection using the Advanced Search tool, and the Citation Report tool was used to retrieve publications and their annual number of citations. The publication and citation windows were 2014–2017 and 2019–2022, respectively, except in the topic of cancer, where I retrieved the publications from 2017 to have fewer than 200,000 publications, which is the limit of publications shown by the WoS. The search in the topic of composite materials excluded publications in the other four topics recorded in Table 2. To facilitate searching, global searches only included the 75 countries with the highest numbers of publications and citations (Clarivate's InCites), plus Latvia and Malta to complete the 27 countries of the EU. Reducing the world to 77 countries had no effect on the results because the papers included in the real global list of papers but omitted from this study did not receive a sufficient number of citations to affect the rank analyses of the countries included in this study.

Searches were restricted to "articles" (DT=Articles). Furthermore, in biomedical topics I manually deleted papers whose titles stated that they were statistical reports or treatment guidelines (eventually, I also used the Abstract). These papers have a significant effect on the $Rk$-index if they are among around the 50 world most-cited papers.

## 3. Results

### 3.1. Technological topics

I studied the research topics of graphene, semiconductors, solar cells/photovoltaics, lithium batteries, and composite materials. The total number of publications ($N$) and the $Rk$-indices are presented in Table 2. Concerning domestic publications, the two most important contributors to the progress of knowledge are the USA and China. The USA ($Rk$-indices ranging from 26 to 32) is notably ahead of China ($Rk$-indices ranging from 13 to 23), except in graphene, where the difference is small (25.7 versus 23.2). In the topic of solar cells/photovoltaics, South Korea, the UK, and Switzerland are the second, third, and fourth-ranked countries, respectively, all ahead of the EU. The EU ranks third ($Rk$-indices ranging from 7 to 13) in the other four topics. Remarkably, Singapore, a

| Country | Graphene | | Semiconductors | | Solar cells/ photovoltaics[*] | | Lithium batteries[*] | | Composite materials | |
|---|---|---|---|---|---|---|---|---|---|---|
| | N | Rk | N | Rk | N | Rk | N | Rk | N | Rk |
| USA(D) | 4677 | 25.68 | 6019 | 28.29 | 5505 | 25.05 | 3376 | 32.22 | 4011 | 28.09 |
| USA(C) | 7303 | 29.09 | 5483 | 25.44 | 5108 | 19.78 | 2921 | 23.20 | 3392 | 27.20 |
| China(D) | 32745 | 23.18 | 12183 | 15.32 | 12883 | 13.10 | 12577 | 20.63 | 11731 | 19.58 |
| China(C) | 8868 | 26.86 | 4312 | 20.97 | 4571 | 15.23 | 3814 | 23.23 | 3271 | 26.22 |
| EU(D) | 5854 | 11.05 | 7452 | 13.13 | 8220 | 7.29 | 2972 | 9.48 | 8831 | 13.13 |
| EU(C) | 5869 | 13.01 | 6564 | 17.23 | 5898 | 16.67 | 1716 | 12.61 | 4367 | 17.79 |
| South Korea(D) | 4704 | 3.61 | 2672 | 3.78 | 4208 | 17.37 | 2243 | 4.76 | 1523 | 3.94 |
| South Korea(C) | 2272 | 6.97 | 1349 | 12.18 | 1663 | 9.01 | 932 | 4.16 | 680 | 5.32 |
| UK(D) | 667 | 1.42 | 811 | 8.18 | 894 | 12.64 | 238 | 2.07 | 759 | 3.51 |
| UK(C) | 2097 | 9.01 | 2005 | 16.47 | 1997 | 17.78 | 581 | 3.18 | 1404 | 8.87 |
| Japan(D) | 1318 | 1.74 | 3171 | 5.33 | 2751 | 4.01 | 1344 | 4.65 | 994 | 6.95 |
| Japan(C) | 1911 | 6.13 | 1533 | 10.74 | 1438 | 9.82 | 573 | 4.04 | 734 | 7.04 |
| Singapore(D) | 686 | 6.99 | 275 | 1.96 | 365 | 1.92 | 317 | 3.04 | 194 | 1.79 |
| Singapore(C) | 1525 | 7.94 | 775 | 5.47 | 758 | 4.82 | 626 | 8.53 | 344 | 8.77 |
| Germany(D) | 844 | 1.09 | 1752 | 3.33 | 1841 | 2.44 | 1142 | 5.56 | 1322 | 2.84 |
| Germany(C) | 2191 | 12.49 | 2902 | 11.31 | 2331 | 8.92 | 904 | 13.39 | 1492 | 11.84 |
| Switzerland(D) | 136 | 0.46 | 281 | 1.95 | 347 | 8.12 | 100 | 0.60 | 214 | 1.72 |
| Switzerland(C) | 460 | 2.29 | 696 | 8.24 | 953 | 19.68 | 122 | 0.81 | 457 | 4.20 |
| Canada(D) | 513 | 1.84 | 535 | 1.28 | 555 | 2.41 | 346 | 3.90 | 642 | 2.23 |
| Canada(C) | 858 | 4.05 | 698 | 5.71 | 759 | 8.12 | 481 | 6.29 | 636 | 4.96 |
| Australia(D) | 552 | 3.83 | 298 | 1.16 | 804 | 2.24 | 236 | 2.18 | 466 | 1.82 |
| Australia(C) | 1798 | 6.59 | 854 | 4.45 | 1220 | 3.59 | 811 | 4.54 | 848 | 5.04 |
| France(D) | 413 | 0.85 | 800 | 1.54 | 681 | 0.77 | 354 | 1.71 | 1046 | 5.33 |
| France(C) | 1265 | 3.24 | 1963 | 5.79 | 1441 | 4.46 | 489 | 3.85 | 1288 | 2.33 |
| India(D) | 3798 | 0.94 | 3059 | 2.05 | 3141 | 1.57 | 661 | 0.75 | 2693 | 3.44 |
| India(C) | 1688 | 1.36 | 912 | 2.00 | 1308 | 1.56 | 381 | 1.00 | 650 | 5.10 |
| Italy(D) | 666 | 0.40 | 657 | 1.90 | 1162 | 2.06 | 187 | 0.41 | 997 | 2.14 |
| Italy(C) | 1062 | 3.07 | 1098 | 5.28 | 1064 | 7.40 | 262 | 1.44 | 855 | 4.20 |
| Spain(D) | 657 | 1.06 | 498 | 1.40 | 797 | 1.42 | 179 | 0.51 | 648 | 2.01 |
| Spain(C) | 1270 | 4.88 | 972 | 4.89 | 1204 | 4.23 | 290 | 2.49 | 798 | 2.94 |
| Netherlands(D) | 133 | 0.65 | 201 | 1.03 | 305 | 0.68 | 26 | 0.44 | 172 | 0.85 |
| Netherlands(C) | 379 | 4.19 | 540 | 4.65 | 563 | 4.12 | 96 | 0.77 | 331 | 2.49 |
| Sweden(D) | 173 | 0.26 | 171 | 0.39 | 236 | 0.85 | 147 | 0.99 | 182 | 0.68 |
| Sweden(C) | 612 | 2.08 | 592 | 3.35 | 847 | 4.79 | 196 | 1.68 | 342 | 1.37 |
| Brazil(D) | 499 | 0.25 | 391 | 0.25 | 181 | 0.67 | 55 | < 0.20 | 1001 | 1.59 |
| Brazil(C) | 547 | 0.56 | 385 | 1.21 | 402 | 3.58 | 38 | < 0.20 | 578 | 1.71 |
| Denmark(D) | 114 | 0.20 | 90 | 0.99 | 126 | 0.54 | 34 | < 0.20 | 124 | 0.70 |
| Denmark(C) | 298 | 1.17 | 277 | 3.07 | 310 | 1.09 | 92 | 0.41 | 167 | 0.84 |
| Greece(D) | 184 | 0.21 | 147 | 0.29 | 216 | 0.32 | 12 | < 0.20 | 195 | 0.85 |
| Greece(C) | 192 | 0.33 | 189 | 0.91 | 246 | 0.65 | 23 | < 0.20 | 178 | 1.18 |
| Austria(D) | 43 | < 0.20 | 127 | 0.37 | 105 | 0.65 | 41 | 0.25 | 100 | 0.36 |
| Austria-Coll | 261 | 0.34 | 357 | 3.96 | 226 | 1.39 | 96 | 0.82 | 235 | 3.39 |
| Belgium(D) | 111 | < 0.20 | 173 | 0.30 | 330 | 0.23 | 74 | 0.25 | 170 | 0.74 |
| Belgium(C) | 420 | 0.69 | 527 | 3.57 | 276 | 0.71 | 159 | 2.44 | 312 | 4.31 |
| Poland(D) | 476 | 0.34 | 484 | 0.34 | 281 | < 0.20 | 124 | < 0.20 | 1016 | 0.77 |
| Poland(C) | 348 | 0.64 | 575 | 1.11 | 288 | 0.70 | 108 | 0.83 | 356 | 1.52 |
| Portugal(D) | 100 | < 0.20 | 86 | 0.30 | 133 | 0.34 | 24 | < 0.20 | 265 | 0.48 |
| Portugal(C) | 213 | < 0.20 | 215 | 0.42 | 206 | 0.47 | 62 | 0.44 | 331 | 1.60 |
| Finland(D) | 95 | 0.21 | 112 | 0.24 | 128 | 0.27 | 24 | < 0.20 | 111 | 0.36 |
| Finland(C) | 301 | 0.79 | 302 | 0.80 | 228 | 0.51 | 26 | < 0.20 | 212 | 0.62 |
| Norway(D) | 17 | < 0.20 | 27 | < 0.20 | 87 | < 0.20 | 29 | 0.41 | 53 | < 0.20 |
| Norway(C) | 79 | < 0.20 | 100 | 0.32 | 172 | 0.29 | 40 | < 0.20 | 96 | 0.34 |

Abbreviations: (D) domestic; (C) collaborations. Publication window: 2014–2017, citation window: 2019–2022.
[*] Data for USA, China, Japan, South Korea, Germany and Singapore reproduced from a previous publication (Rodríguez-Navarro & Brito, 2024b)

very small country, has the fourth highest *Rk*-index for graphene. Considering all five technologies together, South Korea, the UK, and Japan are the next countries behind the EU, while other countries make much lower or negligible contributions. Japan, poorly evaluated with common citation metrics (Bornmann & Leydesdorff, 2013; Pendlebury, 2020), is redeemed by the *Rk*-index and is ahead of other advanced countries such as Germany, Switzerland, Canada, and Australia. Table 3, which summarizes the results of Table 2 considering three thresholds for the *Rk*-index, shows that the USA is the unequivocal leader, while the EU significantly lags behind China.

**Table 3.** Countries' number of topics exceeding three *Rk*-index thresholds in the five technological topics studied here

| *Rk* > 5 | *Rk* > 15 | *Rk* > 25 |
|---|---|---|
| **Domestic** | | |
| USA (5) | USA (5) | USA (5) |
| China (5) | China (4) | |
| EU (5) | South Korea (1) | |
| UK (2) | | |
| Japan (2) | | |
| South Korea (1) | | |
| Switzerland (1) | | |
| Singapore (1) | | |
| Germany (1) | | |
| France (1) | | |
| **International Collaborations** | | |
| USA (5) | USA (5) | USA (3) |
| China (5) | China (5) | |
| EU (5) | EU (3) | |
| Germany (5) | UK (2) | |
| UK (4) | Switzerland (1) | |
| Singapore (4) | | |
| South Korea (4) | | |
| Japan (4) | | |
| Switzerland (2) | | |
| Australia (2) | | |
| Switzerland (2) | | |
| Italy (2) | | |
| France (1) | | |
| Canada (1) | | |

Summary of data in Table 2. In brackets, number of topics

Regarding international collaborations, in the USA, collaborations decrease the *Rk*-index observed for domestic papers, with the exception of graphene. In all other countries, with very few exceptions, the *Rk*-index is higher for collaborative papers than for domestic papers (Tables 2 and 3). Notably, in South Korea, which stands out as a leader in solar cells/photovoltaics, the *Rk*-index for domestic publications is markedly higher than for collaborative publications (17.4 versus 9.0, respectively). In some countries with low values for domestic *Rk*-indices (0.5–2), the ratios between collaborative and domestic *Rk*-indices range from around 2 to 4, with some cases exceeding 11, as in Germany for graphene.

*3.2. Biomedical topics*

I studied the research topics of cancer, stem cells, immunity, inflammation, and Alzheimer's and Parkinson's diseases. Table 4 presents the number of papers and *Rk*-indices for these topics. Regarding domestic publications, apart from the USA, no other country has a single *Rk*-index above 15. In the USA, the *Rk*-indices exceed 15 in all five topics, and in four topics, they surpass 25. The EU, with *Rk*-indices between 10 and 14, outperforms China in all topics except immunity. Excluding the USA, the EU, and China, only the UK for inflammation and Japan for immunity have *Rk*-indices above 5.0.

This scenario of low domestic *Rk*-indices across countries changes dramatically in international collaborations, where it appears that most countries contribute significantly to the progress of knowledge (*Rk*-index above 5). Among the 130 *Rk*-indices calculated, only 33 are lower than 5, and out of the 25 countries studied, only Greece has no *Rk*-indices above 5. Furthermore, there are 43 *Rk*-indices above 15 and 16 above 25. These high indices imply that in most cases the ratio between international collaborative and domestic indices are very high. In the topic of cancer, the ratios between collaborative and domestic *Rk*-indices are above 10 in 20 countries, and in nine countries (Spain, Sweden, Brazil, Norway, Belgium, Poland, Finland, Portugal, and Greece), this ratio is above 30. In other topics, the ratios are not as high but still remarkably high, with only four countries and the EU having no ratios above 10. In contrast to this pattern, the USA's ratios are slightly above 1.0, except in the topic of Alzheimer's and Parkinson's diseases, where it reaches 2.0. The EU and China have higher ratios than the USA but still relatively low.

*3.3. The USA's dominance in international collaborative papers*

Considering that the ratios between the *Rk*-indices of collaborative and domestic papers are remarkably lower for the USA than in most other cases, especially in biomedical

**Table 4.** Number of publications and *Rk*-index in biomedical topics

| Country | Cancer | | Stem cells | | Immunity | | Inflammation | | Alzheimer/ Parkinson | |
|---|---|---|---|---|---|---|---|---|---|---|
| | N | Rk | N | Rk | N | Rk | N | Rk | N | Rk |
| USA(D) | 17501 | 28.40 | 16464 | 33.10 | 7963 | 25.58 | 18694 | 28.08 | 7202 | 19.07 |
| USA(C) | 12801 | 34.79 | 13083 | 32.20 | 6833 | 31.47 | 14711 | 33.48 | 5754 | 38.47 |
| EU(D) | 16065 | 13.83 | 13931 | 9.66 | 6694 | 10.23 | 20319 | 10.70 | 8097 | 13.67 |
| EU(C) | 9610 | 30.73 | 9843 | 28.20 | 5542 | 25.27 | 12093 | 33.32 | 5647 | 36.29 |
| China(D) | 24114 | 9.28 | 14713 | 3.64 | 6752 | 11.71 | 18423 | 10.18 | 4486 | 4.31 |
| China(C) | 5770 | 25.55 | 5585 | 21.31 | 2462 | 10.96 | 5332 | 13.62 | 1725 | 7.50 |
| UK(D) | 1901 | 3.46 | 1769 | 3.36 | 894 | 3.77 | 2176 | 8.42 | 1021 | 3.59 |
| UK(C) | 4045 | 24.89 | 4006 | 16.43 | 2639 | 17.69 | 4780 | 30.34 | 2808 | 28.86 |
| Japan(D) | 5389 | 2.78 | 4742 | 2.99 | 1494 | 5.41 | 5092 | 3.08 | 1510 | 2.75 |
| Japan(C) | 1674 | 23.20 | 2386 | 12.92 | 977 | 9.13 | 2020 | 9.04 | 699 | 7.83 |
| Germany(D) | 2774 | 1.55 | 2928 | 3.08 | 1179 | 3.62 | 3596 | 3.07 | 1154 | 4.04 |
| Germany(C) | 3533 | 24.47 | 4223 | 21.05 | 2087 | 17.51 | 4838 | 21.09 | 2151 | 28.79 |
| Canada(D) | 1679 | 1.41 | 1230 | 0.96 | 734 | 1.97 | 1765 | 3.06 | 790 | 4.04 |
| Canada(C) | 2516 | 24.49 | 1947 | 16.55 | 1086 | 9.64 | 2571 | 10.91 | 1311 | 14.46 |
| France(D) | 1891 | 3.10 | 1534 | 1.02 | 789 | 1.47 | 1642 | 2.33 | 858 | 1.93 |
| France(C) | 2393 | 27.83 | 2346 | 13.07 | 1540 | 15.72 | 2486 | 14.35 | 1375 | 19.57 |
| South Korea(D) | 3908 | 2.05 | 3282 | 2.61 | 1090 | 1.13 | 4713 | 1.69 | 1278 | 1.95 |
| South Korea(C) | 1248 | 23.89 | 1342 | 9.54 | 505 | 4.90 | 1291 | 2.98 | 586 | 5.01 |
| Italy(D) | 2595 | 1.12 | 2179 | 1.91 | 788 | 1.71 | 2879 | 1.85 | 1356 | 2.22 |
| Italy(C) | 2635 | 24.37 | 2381 | 11.43 | 901 | 7.03 | 2735 | 16.36 | 1487 | 12.01 |
| Netherlands(D) | 1049 | 1.45 | 127 | 0.24 | 472 | 1.51 | 1241 | 2.52 | 443 | 2.08 |
| Netherlands(C) | 1747 | 25.23 | 353 | 1.74 | 1063 | 12.34 | 2413 | 15.89 | 907 | 24.13 |
| Australia(D) | 1276 | 1.05 | 882 | 0.55 | 600 | 0.97 | 1505 | 1.80 | 661 | 2.00 |
| Australia(C) | 1881 | 24.87 | 1583 | 12.82 | 1085 | 9.62 | 2292 | 11.24 | 1013 | 21.38 |
| Spain(D) | 1131 | 0.70 | 1209 | 2.22 | 504 | 0.87 | 1767 | 1.16 | 941 | 1.34 |
| Spain(C) | 1726 | 23.38 | 1755 | 17.54 | 761 | 7.05 | 1738 | 11.68 | 1103 | 12.87 |
| Switzerland(D) | 402 | 1.03 | 454 | 1.55 | 219 | 1.03 | 504 | 1.57 | 143 | 0.27 |
| Switzerland(C) | 1415 | 17.33 | 1469 | 11.72 | 821 | 12.08 | 1681 | 9.71 | 737 | 14.89 |
| India(D) | 2233 | 0.61 | 1127 | 0.86 | 963 | 0.57 | 2056 | 0.71 | 889 | 1.94 |
| India(C) | 931 | 13.13 | 594 | 1.84 | 416 | 1.38 | 701 | 1.25 | 307 | 2.18 |
| Sweden(D) | 603 | 0.49 | 461 | 1.17 | 234 | 0.75 | 892 | 0.69 | 358 | 1.01 |
| Sweden(C) | 1313 | 17.44 | 1163 | 12.06 | 715 | 7.00 | 1749 | 16.83 | 1239 | 25.50 |
| Brazil(D) | 1106 | 0.45 | 967 | 0.48 | 711 | 0.52 | 2930 | 0.91 | 562 | 1.33 |
| Brazil(C) | 721 | 15.30 | 654 | 5.33 | 575 | 3.85 | 1456 | 4.10 | 405 | 3.25 |
| Norway(D) | 290 | 0.36 | 702 | 1.61 | 103 | 0.23 | 343 | 0.37 | 88 | 0.83 |
| Norway(C) | 608 | 11.61 | 1671 | 14.84 | 272 | 3.11 | 686 | 4.08 | 263 | 2.37 |
| Austria(D) | 244 | 0.53 | 201 | 0.80 | 100 | 0.51 | 392 | 0.85 | 112 | 0.47 |
| Austria(C) | 770 | 10.51 | 749 | 8.66 | 358 | 7.60 | 947 | 9.75 | 303 | 6.96 |
| Denmark(D) | 527 | 0.40 | 199 | 0.29 | 176 | 0.46 | 677 | 0.56 | 168 | 1.41 |
| Denmark(C) | 879 | 11.21 | 595 | 5.60 | 458 | 5.88 | 1107 | 12.39 | 329 | 4.70 |
| Belgium(D) | 330 | 0.54 | 282 | 0.40 | 225 | 0.71 | 442 | 0.69 | 141 | 0.42 |
| Belgium(C) | 1035 | 21.61 | 906 | 5.63 | 543 | 5.63 | 1129 | 12.70 | 525 | 5.29 |
| Poland(D) | 881 | 0.37 | 570 | 0.25 | 367 | 0.35 | 1251 | 0.37 | 247 | 1.27 |
| Poland(C) | 609 | 16.56 | 470 | 2.95 | 187 | 4.55 | 516 | 1.97 | 282 | 1.82 |
| Singapore(D) | 247 | 0.42 | 328 | 0.98 | 75 | 0.37 | 159 | < 0.20 | 108 | 0.50 |
| Singapore(C) | 627 | 4.59 | 879 | 7.31 | 280 | 4.58 | 508 | 2.50 | 211 | 1.31 |
| Finland(D) | 220 | 0.21 | 177 | < 0.20 | 112 | 0.22 | 399 | 0.39 | 122 | 0.88 |
| Finland(C) | 529 | 11.79 | 412 | 1.90 | 249 | 4.16 | 613 | 7.22 | 329 | 4.69 |
| Portugal(D) | 264 | 0.28 | 242 | 0.38 | 84 | 0.23 | 281 | 0.37 | 176 | 0.43 |
| Portugal(C) | 448 | 14.87 | 410 | 1.88 | 196 | 2.98 | 343 | 1.39 | 300 | 2.11 |
| Greece(D) | 246 | 0.12 | 147 | 0.21 | 78 | < 0.2 | 358 | < 0.20 | 109 | 0.27 |
| Greece(C) | 400 | 4.33 | 254 | 3.15 | 109 | 4.07 | 460 | 4.38 | 230 | 3.27 |

Abbreviations: (D) domestic; (C) collaborations. Publication window: 2014–2017, except in cancer where it was 2017, citation window: 2019–2022.

topics, a reasonable hypothesis was that the USA is playing a central role in these collaborations. To test this hypothesis, I focused on studying of the EU and China collaborations with the USA in biomedical topics. First, I recalculated the *Rk*-indices for collaborative papers excluding those with US participation. The results show that, upon excluding the USA, the *Rk*-indices are drastically reduced (Table 5, which also includes the data in Table 4 for comparison). Considering the leadership of China and the EU, it can be concluded that these reductions likely occur in all countries. For example, for cancer in Spain, which is a country with a large number of publications, the exclusion of the USA in collaborative papers reduces the *Rk*-index from 23.4 (Table 4) to 4.3 (data not shown). Fig. 1 visually summarizes the results in the topics of cancer and stem cells.

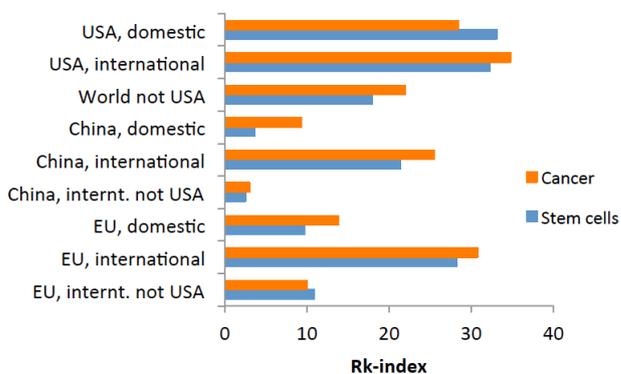

**Fig. 1.** *Rk*-index of publications from the USA, China, and the EU. Papers are sorted into three categories: domestic, international collaborations, and international collaborations in China and the EU excluding the participation of the USA. Publication window: 2014–2017, citation window: 2019–2022

Next, I made similar recalculations for technological topics. The results were similar although the differences were less pronounced. Interestingly, the exclusion of US participation decreased the *Rk*-indices more in China than in the EU (Table 5).

These results indicate that the high *Rk*-indices of international collaborative papers are high because of US collaboration. Considering also the superiority of the USA in domestic papers, the whole suggests a clear dominance of the USA in world research. To further investigate this possibility, I calculated the *Rk*-indices of all world papers without US participation to compare them with those of the US domestic papers. In other words, I compared publications solely from the USA with world publications excluding the USA (Table 6). In biomedical topics, with the exception of the Alzheimer's/Parkinson's topic, the *Rk*-indices are higher for US domestic papers than for world papers excluding the USA, indicating that the US contribution to pushing the boundaries of knowledge is higher than that of all the rest of the countries. In

**Table 5.** The role of the USA in international collaborative publications of the EU and China analyzed by the *Rk*-index

| | EU | | | | China | | | |
|---|---|---|---|---|---|---|---|---|
| | All collaborations | | Collaborations not including USA | | All collaborations | | Collaborations not including USA | |
| | *N* | *Rk* | *N* | *Rk* | *N* | *Rk* | *N* | *Rk* |
| **Biomedical topics** | | | | | | | | |
| Cancer | 9610 | 30.73 | 5029 | 10.03 | 5770 | 25.55 | 1948 | 3.04 |
| Stem Cells | 9843 | 28.20 | 4974 | 10.88 | 5585 | 21.31 | 1863 | 2.45 |
| Immunity | 5542 | 25.27 | 2956 | 9.11 | 2462 | 10.96 | 924 | 1.79 |
| Inflammation | 12093 | 33.32 | 6412 | 9.21 | 5332 | 13.62 | 1777 | 3.84 |
| Alzheimer/Parkinson | 5647 | 36.29 | 3118 | 11.79 | 1725 | 7.50 | 612 | 3.01 |
| **Technological topics** | | | | | | | | |
| Graphene | 5869 | 13.01 | 4495 | 10.22 | 8868 | 26.86 | 5161 | 12.07 |
| Semiconductors | 6564 | 17.23 | 4761 | 13.20 | 4312 | 20.97 | 2473 | 13.01 |
| Solar cells/photovoltaics | 5898 | 16.67 | 3956 | 15.09 | 4571 | 15.23 | 2429 | 8.07 |
| Lithium batteries | 1716 | 12.61 | 1309 | 9.12 | 3814 | 23.23 | 2105 | 8.61 |
| Composite materials | 4367 | 17.79 | 3516 | 10.93 | 3271 | 26.22 | 1995 | 15.66 |

"All collaborations" columns duplicate data from Tables 4 and 2

**Table 6.** Number and *Rk*-index of publications from USA only, the world excluding USA, China only, and the world excluding China

| | Domestic USA | | World not USA | | Domestic China | | World not China | |
|---|---|---|---|---|---|---|---|---|
| | N | Rk-index | N | Rk-index | N | Rk-index | N | Rk-index |
| **Biomedical topics** | | | | | | | | |
| Cancer | 17501 | 28.4 | 74282 | 21.9 | 24114 | 9.3 | 74698 | 35.3 |
| Stem Cells | 16464 | 33.1 | 57166 | 18.0 | 7663 | 3.7 | 66400 | 39.5 |
| Immunity | 7963 | 25.6 | 27867 | 24.2 | 6752 | 11.7 | 33451 | 38.0 |
| Inflammation | 18694 | 28.1 | 80814 | 24.0 | 18423 | 10.2 | 90470 | 39.5 |
| Alsheimer/Parkinson | 7202 | 19.1 | 26717 | 22.4 | 4486 | 4.3 | 33462 | 39.5 |
| **Technological topics** | | | | | | | | |
| Graphene | 4677 | 25.7 | 70682 | 29.2 | 32745 | 23.2 | 41057 | 34.5 |
| Semiconductors | 6019 | 28.3 | 46681 | 33.4 | 12183 | 15.3 | 41783 | 36.2 |
| Solar cells/photovoltaics | 5505 | 25.1 | 43571 | 36.2 | 12883 | 13.1 | 43934 | 39.3 |
| Lithium batteries | 3376 | 32.2 | 26003 | 27.9 | 12577 | 20.6 | 15908 | 36.2 |
| Composite materials | 4011 | 28.1 | 42676 | 30.2 | 11731 | 19.6 | 35075 | 33.3 |

Publication window 2014–2017, citation window 2019–2022. "Domestic" columns duplicate data from Tables 4 and 2

technological topics, with the exception of lithium batteries, the *Rk*-indices are lower for US domestic papers than for the world without the USA, but the differences are low. The highest difference is in solar cells/photovoltaics, where the ratio is 1.4. It is worth noting that the described differences occur with a number of USA domestic papers that is much lower than the number of the world papers without the USA—3–4 times lower in biomedical topics and 8–15 times lower in technological topics.

I also used this approach to further address the controversy about the USA or China leading the worldwide progress of knowledge. For this purpose, I calculated the *Rk*-indices of world publications without China's participation. To better interpret the results, it is worth noting that the *Rk*-index of all world papers is 39.5. If a country has a low global contribution, the elimination of the papers in which that country appears from the global papers will decrease little or nothing the value of 39.5. Under this perspective, the conclusion that can be drawn from the results in Table 6 is that overall in 10 topics studied here, the contribution of China is lower than the contribution of the USA. In biomedical topics, the difference is notable; excluding the USA, the resulting *Rk*-indices are below 25, and excluding China, the resulting *Rk*-indices are above 33.

## 4. Discussion

### 4.1. Disclosures with the Rk-index

The *Rk*-index is specifically designed to disclose the contribution of countries or institutions to pushing the boundaries of knowledge (Rodríguez-Navarro & Brito, 2024b) and is based on the global ranks of the 10 most cited papers of each country or institution (Table 1). The analysis of these ranks intuitively shows the advantage of some countries over others, and the *Rk*-index condenses into a single nonparametric indicator the information provided by the ranks. In comparison with other well-known nonparametric indicators such as the number of top 1% most cited papers, the *Rk*-index is much more stringent, at the level of the top 0.01% most cited papers, but it can be calculated even for countries that have never published a paper at that level of stringency. Furthermore, it is insensitive to any possible deviation of the most cited papers from the ideal model in rank distribution (Rodríguez-Navarro, 2024).

It is worth noting, however, that the *Rk*-index only provides specific information about the contribution of countries to pushing the boundaries of knowledge. A complete description of a research system should also consider incremental developments in scientific topics, which constitute the major part of countries' research, and technical incremental innovations, which receive few or no citations (Rodríguez-Navarro, 2024);Rodríguez-Navarro, 2021 #1996].

The main conclusion that can be drawn from the results of this study is that the USA dominates the global contribution to pushing the boundaries of knowledge. This conclusion is clear considering domestic papers (Tables 2–4), and it is even clearer considering international collaborative papers, but only if the results are properly analyzed. The contribution of international collaborative papers to global research is high (Wagner et al., 2015), and any error on the calculation of the distribution of credits between the participating countries may lead to significant mistakes in research assessments.

In the 260 cases studied here (Tables 2 and 4), the ratio between the numbers of domestic and international collaborative papers ranges from around 4.0 (e.g., India in composite materials) to around 0.3 (e.g., the UK or Australia in graphene). On average, 43.6% of publications are international collaborations. The general pattern observed in many countries is that international collaborative papers have higher $Rk$-indices than domestic papers, which could be interpreted as evidence of the collaborative papers' dominance in the progress of knowledge. This interpretation has been questioned (Olechnicka et al., 2019; Wagner et al., 2019; Zanotto et al., 2016) and the present results demonstrate that international collaborations reveal the dominance of the USA and how this dominance may be improperly extended to other countries. For instance, in the topic of inflammation, the EU's $Rk$-index in international collaborative papers is 33.3, which is a high index, but it is reduced to 9.2 if collaborations with the USA are eliminated (Table 5).

The USA's dominance described in this study across 10 research topics contradicts the conclusions of other studies. For example, it contradicts the *ASPI's Technology Tracker* (Gaida et al., 2023), which evaluates based on the number of top 10% most cited papers and the *h*-index, and the *National Institute of Science and Technology Policy* in Tokyo (National Institute of Science and Technology, 2022), which evaluates using the number of top 1% most cited papers. Both reports place China ahead of the USA. Although the studied periods are different, 2014–2017 in this study, and 2018–2022 and 2018–2020, respectively, in the other two studies, it is unlikely that this is the cause of the differences. Differences in the methods of evaluations can better explain the differences. Firstly, the numbers of top 10% and 1% most cited papers are not suitable indicators for breakthrough papers, which are 100 or 1000 times less frequent. Secondly, the fractional counting used in these reports is also does not take into account the US dominance in collaborative papers.

Apart from the cases of the USA and China, the present results show the subordinate position of the EU concerning to the USA in both technological and biomedical topics, and its subordination concerning to China in technological topics. In biomedical topics, the EU is better positioned than China, but its most important success depends on its collaboration with the USA (Table 5 and Fig. 1).

The main conclusion regarding countries other than the USA, China, and the EU is that, overall, their contribution to pushing the boundaries of knowledge is low, but there are exceptions, such as the case of South Korea in solar cells/photovoltaics. Apart from confirming that Japan is not a scientifically undeveloped country (Rodríguez-Navarro & Brito, 2024a), the use of the *Rk*-index strongly suggests that the higher position of Australia and Canada compared to South Korea in country rankings (e.g., European Commission, 2022) may not be accurate (Tables 2 and 4). It also indicates that, despite significant scientific growth in BRICS countries (Bornmann et al., 2015), two major BRICS countries, India and Brazil, still lag considerably behind the leading countries in research (Tables 2 and 4).

Although I only investigated 10 research topics, five in technological and five in biomedical fields, these topics are currently of extraordinary importance and are investigated in all advanced countries, although small countries may focused on only some of them. Therefore, the present results can be considered representative of the general contribution of countries to pushing the boundaries of knowledge. Further studies could correct some limitations of the *Rk*-index (Rodríguez-Navarro & Brito, 2024b) and improve the use of the ranks of the most cited papers (Table 1) for evaluating countries on the basis of their contribution to pushing the boundaries of knowledge, but it can be ruled out that these improvements affect to the general picture described in this study.

### 4.1. Lessons for research policy

The research success of countries with a sufficient economic level is a direct consequence of their research policies. The success of China's research, as shown in Table 2 and coincident with other studies (Adams et al., 2023), is an obvious result of its research policy (Benner et al., 2012) because China's GNI per capita is not high.

Similarly, the lower performance of the EU in comparison with the USA should be the result of its research policy. The USA's GNI per capita is higher than that of the EU (https://data.worldbank.org/indicator/NY.GNP.PCAP.PP.CD?locations=US, accessed on 2/21/2024), but it is unlikely that the economy is the cause of the difference. More likely, the research policy makes the difference. Since the EU's coining of the catchword "*European Paradox*" in 1994 (Albarrán et al., 2010) and its replacement by "*Europe is a global scientific powerhouse*" in 2017 (Rodríguez-Navarro & Brito, 2020), the European Commission (EC) has based its research policy on the excellence of the EU's research. The present results demonstrate that the basis of the EU's research policy, the EU's research success, is not true and that the EU needs a more realistic research policy.

*4.2. Further implications*

One of the most interesting analyses of the research policies of countries is the evolution of their contribution to the progress of knowledge. So far, this has been addressed using the number of Nobel Prizes (Gros, 2018). This approach has the inconvenience of the infrequency of Nobel Prizes, which restricts its use to very few countries in order to have results with reasonable statistical significance. Further analyses based on the *Rk*-index instead of the number of Nobel Prizes could provide reliable data to implement the most convenient research policy to achieve important discoveries.

**CONFLICT OF INTEREST**

The author declares that there is no conflict of interest.